\def\nk{n_{\rm b}}

\def\Pb{P_{\rm b}}

\def\rfr#1{Equation\,(\ref{#1})}

\def\dert#1#2{\frac{{{\textrm{d}}}{#1}}{{{\textrm{d}}}{#2}}}

\def\eqi{\begin{equation}}
\def\eqf{\end{equation}}
\def\eqia{\begin{eqnarray}}
\def\eqfa{\end{eqnarray}}

\def\rp#1#2{{#1\over#2}}
\def\lb#1{\label{#1}}

\def\bds#1{\boldsymbol{#1}}

\def\ton#1{\left(#1\right)}
\def\qua#1{\left[#1\right]}

\def\ang#1{\left\langle #1\right\rangle}
\documentclass[linenumbers]{aastex}

\usepackage{morefloats}
\usepackage[title]{appendix}
\usepackage{hyperref,textcomp}
\usepackage{booktabs}
\usepackage[table,xcdraw]{xcolor}
\usepackage{multirow}
\usepackage{rotating,tabularx}
\usepackage{float}
\usepackage{enumerate}
\usepackage{rotating}
\usepackage[polutonikogreek,english]{babel}
\usepackage{amsmath,starfont,textgreek,w-greek,wasysym}
\usepackage[flushleft]{threeparttable}
\usepackage{amsthm}
\usepackage{amscd,lineno}
\usepackage{amssymb,dsfont}
\usepackage{graphicx,epsfig}
\usepackage[utf8]{inputenc}
\usepackage[T1]{fontenc}
\usepackage{txfonts}
\usepackage{xr-hyper}

\RequirePackage{color}

\newcommand{\emaila}{lorenzo.iorio@libero.it}

\allowdisplaybreaks[1]

\begin{document}

\title{Why the mean anomaly at epoch is not used in tests of non-Newtonian gravity?}

\shortauthors{L. Iorio}

\author{Lorenzo Iorio\altaffilmark{1} }
\affil{Ministero dell'Istruzione, dell'Universit\`{a} e della Ricerca
(M.I.U.R.)
\\ Viale Unit\`{a} di Italia 68, I-70125, Bari (BA),
Italy}

\email{\emaila}

\begin{abstract}
The mean anomaly at epoch $\eta$ is one of the standard six Keplerian orbital elements in terms of which the motion of the two-body problem is parameterized. Along with the argument of pericenter $\omega$, $\eta$ experiences long-term rates of change induced, among other things, by general relativity and several modified models of gravity. Thus, in principle, it may be fruitfully adopted together with $\omega$ in several tests of post-Newtonian gravity performed with astronomical and astrophysical binary systems. This would allow to enhance the gravitational signature one is interested in and to disentangle some competing disturbing effects acting as sources of systematic bias. Nonetheless, for some reasons unknown to the present author, $\eta$ has never been used so far by astronomers in actual data reductions. This note aims to raise interest in the community about the possible practical use of such an orbital element or, at least, to induce experts in astronomical data processing to explicitly make clear if it is not possible to use $\eta$ for testing gravitational models and, in this case,  why.
\end{abstract}

{
\textit{Unified Astronomy Thesaurus concepts}:   Gravitation\,(661); General relativity\,(641);  Relativistic mechanics\,(1391)
}


\section{Testing post-Newtonian gravity with orbital motions}\lb{sec_1}
Since the explanation in 1915 \citep{1915SPAW...47..831E} of the then anomalous perihelion precession of Mercury \citep{LeVer1859} by Einstein with his general relativity, orbital motions have been used as effective tools to scrutinize  alternative theories of gravitation with respect to the dominant paradigm at the time. Such tests are often presented in terms of secular precessions of the argument of pericentre $\omega$ in two-body systems like Earth's artificial satellites \citep{2010PhRvL.105w1103L,2014PhRvD..89h2002L}, Solar System's planets and asteroids \citep{1968PhRvL..20.1517S,1971AJ.....76..588S,
1972PhRvL..28.1594S,1990grg..conf..313S}, binary pulsars \citep{2004Sci...303.1153L,2021PhRvX..11d1050K} and stars orbiting supermassive black holes \citep{2020A&A...636L...5G,2022A&A...657L..12G}. In all such cases, a perturbative approach can be adopted in predicting the sought effects since they can be thought as consequences of relatively small additional accelerations with respect to the dominant inverse-square Newtonian monopole.
\section{Using the pericentre}\lb{sec_2}
As far as general relativity is concerned, the largest precession, to the first post-Newtonian (pN) level of the order of $\mathcal{O}\ton{c^{-2}}$,  where $c$ is the speed of light in vacuum,  is due to the so-called gravitoelectric part of the spacetime metric depending only on the masses $M_\textrm{A}$ and $M_\textrm{B}$ of the bodies A and B constituting the two-body system at hand  \citep{Sof89,SoffelHan19}. It induces a perturbing acceleration $\bds A$ \citep{Sof89,SoffelHan19} lying in the orbital plane characterized, in general, by both radial $A_R$ and transverse $A_T$ components. When plugged in the machinery of, e.g., the Gauss equations for the variation of the orbital elements \citep{Nobilibook87,Sof89,1991ercm.book.....B,2003ASSL..293.....B,2011rcms.book.....K,SoffelHan19}, it turns out that the semimajor axis $a$, the eccentricity $e$, the inclination $I$ and the longitude of the ascending node $\Omega$ do not undergo net shifts when the average over one full orbital period $\Pb$ is taken, contrary to $\omega$ which, instead, exhibits a secular rate of change\footnote{Here and in the following, the angular brackets $\ang{\ldots}$ denoting the average over $\Pb$ will be omitted for making the notation less cumbersome.}
\citep{Sof89,1991ercm.book.....B,2003ASSL..293.....B,2011rcms.book.....K,SoffelHan19}
\eqi\dot\omega = \rp{3\,\nk\,\mu}{c^2\,a\,\ton{1-e^2}},\lb{omega_GR}\eqf
where $\mu\doteq GM$ is the product of  the Newtonian constant of gravitation $G$ times  the sum of the masses of the binary's constituents $M\doteq M_\textrm{A}+M_\textrm{B}$, and $\nk\doteq 2\uppi/\Pb=\sqrt{\mu/a^3}$ is the Keplerian mean motion.

Long-range modified models of gravity \citep{2007IJGMM..04..115N,Lobo09,2011PhRvD..84b4020H,2012PhR...513....1C,2015Univ....1..199C,2016RPPh...79j6901C,2017PhR...692....1N,universe7080269}, devised in the last decades mainly to explain certain astrophysical and cosmological features\footnote{For a comprehensive overview of such themes, see, e.g., \citet{2016Univ....2...23D} and references therein.} like the phenomenology of Dark Matter \citep{2017ARA&A..55....1F,2018ARA&A..56..435W,2019IJMPA..3430013K} and the accelerated rate of the cosmic expansion \citep{2008ARA&A..46..385F,2018RPPh...81a6902B}, can be used, in principle, to perform local tests as well. Indeed, for most of them, a correction to the Newtonian gravitational potential in the $g_{00}$ term of the spacetime metric can be explicitly worked out. In turn, it yields a perturbing acceleration which in most cases, is entirely radial. As a result, only $\omega$ experiences a net secular advance.
\section{The impact of the orbital perturbations}\lb{sec_3}
Actually, the dynamical evolution of the pericentre of any realistic astronomical and astrophysical system of interest is not determined only by the model of gravity one is looking at, being impacted also by several other competing Newtonian gravitational\footnote{In some cases, like Solar System's asteroids and Earth's artificial satellites, non-gravitational perturbations \citep{Nobilibook87} may play a non-negligible role as well.} effects of different origin (multipole moments of the system's bodies, tides, pulls by bodies external to the system, etc.) which, in the present context, are viewed as sources of systematic bias. Thus, strategies to reduce the systematic errors due to them with respect to the signature of interest must be devised. If the shifts caused by the other dynamical features on the pericentre can be analytically calculated,  it is possible, in principle, to disentangle them from the one searched for by using more than one pericentre precessions provided they are accessible to observations. Such an approach was proposed, for the first time,
by \citet{1990grg..conf..313S} whose goal, at that time, was to separate the Sun-induced pN gravitoelectric perihelion precession from that due to the solar quadrupole mass moment $J_2$ by using the perihelia of other planets or asteroids on highly eccentric orbits. For a general overview of such a strategy, not limited just to our Solar System, see \citet{2019EPJC...79..816I}. Clearly, for each body, it would be better to have at disposal more than one observable affected by the gravity model under consideration; in such a way, one would be less dependent on the presence or not of other bodies whose pericentres should be used as probes.
\section{The potential benefits of the mean anomaly at epoch}\lb{sec_4}
In fact, this is just the case for general relativity and other alternative gravitational theories. Indeed, there is another orbital parameter which is secularly displaced by the 1pN gravitoelectric acceleration and by several modified models of gravity: the mean anomaly at epoch $\eta$.
Let us recall that it is one of the standard six Keplerian orbital elements in terms of which the motion in the two-body problem is parameterized.
The Gauss equation for its rate of change is  \citep{Nobilibook87,Sof89,1991ercm.book.....B,2003ASSL..293.....B,2011rcms.book.....K,SoffelHan19}
\eqi
\dert{\eta}{t} = -\rp{2}{\nk\,a}\,A_R\,\ton{\rp{r}{a}} - \rp{\ton{1-e^2}}{\nk\,a\,e}\qua{-A_R\,\cos f + A_T\,\ton{1+\rp{r}{p}}\,\sin f},\lb{eta_Gauss}
\eqf
where $r$ is the mutual distance between A and B, and $p\doteq a\ton{1-e^2}$ is the semilatus rectum of the Keplerian ellipse.
From \rfr{eta_Gauss} it can be noted that, in principle, $\eta$ is shifted with respect to the unperturbed Keplerian case if an in-plane perturbation is present, as is the case for, e.g., the 1pN gravitoelectric acceleration  \citep{Sof89,SoffelHan19} and for several long-range alternative models of gravitation.
As an example, it can straightforwardly be worked out that the general relativistic Schwarzschild-like acceleration and \rfr{eta_Gauss} yield for $\eta$ the following net rate of change over one orbital revolution
\eqi
\dot\eta =\rp{\mu\,\nk\,\qua{-15 + 6\,\sqrt{1-e^2} + \nu\,\ton{9-7\,\sqrt{1-e^2}}}}{c^2\,a\,\sqrt{1-e^2}},\lb{eta_GR}
\eqf
where
\eqi
\nu\doteq\rp{M_\mathrm{A}\,M_\mathrm{B}}{M^2}.
\eqf
In general, $\eta$ is impacted by the same disturbances as $\omega$. This is an important point since many confuse $\eta$ with the mean anomaly $\mathcal{M}$ whose rate includes also the mean motion $\nk$ which is plagued by the uncertainty with which $\mu$ is known at the time of the data analysis and possible non-gravitational effects in $a$; see \citet{2019EPJC...79..816I} for a discussion of such subtleties.

In principle, using $\eta$ in conjunction with $\omega$ would yield an enhancement of the signal of interest; suffice it to say that, in the case of Mercury orbiting the Sun, \rfr{omega_GR} yields the time-honored $42.98\,\mathrm{arcsec\,cty}^{-1}$, while \rfr{eta_GR} provides us with $-127.986\,\mathrm{arcsec\,cty}^{-1}$. Thus, the question arises: why the mean anomaly at epoch and its secular rate of change have never been used so far in  tests of General Relativity and, more generally, pN gravity? The present author does not know the answer, and hopes that the experts in data reductions  may address such a question providing the community with a clear and unambiguous answer, even if, for some reasons, it were negative about the possible practical use of $\eta$. \textcolor{black}{For example, should $\eta$ be measured as a delay of the secondary's orbital motion as compared to a Keplerian
motion, one would need to know the orbital parameters with a sufficiently high precision. Usually, $a$ can be measured accurately, while getting $e$ with a comparable accuracy may be not so easy.} Such an answer would have a great significance, especially in view of the many systems which are already used or may become adopted in the near future, for testing General Relativity and other modified theories of gravitation like extrasolar planets close to their parent stars, binary pulsars and stars orbiting supermassive black holes. Even from the point of view of history of science, it would be interesting to understand why astronomers missed the much larger general relativistic shift of $\eta$ of Mercury with respect to that of $\omega$.
\bibliography{psrbib}{}

\begin{thebibliography}{33}
\expandafter\ifx\csname natexlab\endcsname\relax\def\natexlab#1{#1}\fi

\bibitem[{Bahamonde \& Said(2021)}]{universe7080269}
Bahamonde S., Said J.~L., 2021, Universe, 7

\bibitem[{{Bertotti}, {Farinella} \& {Vokrouhlick\'{y}}(2003){Bertotti},
  {Farinella}, \& {Vokrouhlick\'{y}}}]{2003ASSL..293.....B}
{Bertotti} B., {Farinella} P., {Vokrouhlick\'{y}} D., 2003, {Physics of the
  Solar System}. Kluwer, Dordrecht

\bibitem[{{Brax}(2018)}]{2018RPPh...81a6902B}
{Brax} P., 2018, Reports on Progress in Physics, 81, 016902

\bibitem[{{Brumberg}(1991)}]{1991ercm.book.....B}
{Brumberg} V.~A., 1991, {Essential Relativistic Celestial Mechanics}. Adam
  Hilger, Bristol

\bibitem[{{Cai} {et~al}\mbox{.}(2016){Cai}, {Capozziello}, {De Laurentis}, \&
  {Saridakis}}]{2016RPPh...79j6901C}
{Cai} Y.-F., {Capozziello} S., {De Laurentis} M., {Saridakis} E.~N., 2016,
  Reports on Progress in Physics, 79, 106901

\bibitem[{{Capozziello} {et~al}\mbox{.}(2015){Capozziello}, {Harko},
  {Koivisto}, {Lobo}, \& {Olmo}}]{2015Univ....1..199C}
{Capozziello} S., {Harko} T., {Koivisto} T., {Lobo} F., {Olmo} G., 2015,
  Universe, 1, 199

\bibitem[{{Clifton} {et~al}\mbox{.}(2012){Clifton}, {Ferreira}, {Padilla}, \&
  {Skordis}}]{2012PhR...513....1C}
{Clifton} T., {Ferreira} P.~G., {Padilla} A., {Skordis} C., 2012, \physrep,
  513, 1

\bibitem[{{Debono} \& {Smoot}(2016)}]{2016Univ....2...23D}
{Debono} I., {Smoot} G.~F., 2016, Universe, 2, 23

\bibitem[{{Einstein}(1915)}]{1915SPAW...47..831E}
{Einstein} A., 1915, Sitzber. Preuss. Akad., 831

\bibitem[{{Freeman}(2017)}]{2017ARA&A..55....1F}
{Freeman} K.~C., 2017, \araa, 55, 1

\bibitem[{{Frieman}, {Turner} \& {Huterer}(2008){Frieman}, {Turner}, \&
  {Huterer}}]{2008ARA&A..46..385F}
{Frieman} J.~A., {Turner} M.~S., {Huterer} D., 2008, \araa, 46, 385

\bibitem[{{Gravity Collaboration} {et~al}\mbox{.}(2022){Gravity Collaboration},
  {Abuter}, {Aimar}, {Amorim}, {Ball}, {Baub{\"o}ck}, {Berger}, {Bonnet},
  {Bourdarot}, {Brandner}, {Cardoso}, {Cl{\'e}net}, {Dallilar}, {Davies}, {de
  Zeeuw}, {Dexter}, {Drescher}, {Eisenhauer}, {F{\"o}rster Schreiber},
  {Foschi}, {Garcia}, {Gao}, {Gendron}, {Genzel}, {Gillessen}, {Habibi},
  {Haubois}, {Hei{\ss}el}, {Henning}, {Hippler}, {Horrobin}, {Jochum}, {Jocou},
  {Kaufer}, {Kervella}, {Lacour}, {Lapeyr{\`e}re}, {Le Bouquin}, {L{\'e}na},
  {Lutz}, {Ott}, {Paumard}, {Perraut}, {Perrin}, {Pfuhl}, {Rabien},
  {Shangguan}, {Shimizu}, {Scheithauer}, {Stadler}, {Stephens}, {Straub},
  {Straubmeier}, {Sturm}, {Tacconi}, {Tristram}, {Vincent}, {von Fellenberg},
  {Widmann}, {Wieprecht}, {Wiezorrek}, {Woillez}, {Yazici}, \&
  {Young}}]{2022A&A...657L..12G}
{Gravity Collaboration} {et~al.}, 2022, \aap, 657, L12

\bibitem[{{Gravity Collaboration} {et~al}\mbox{.}(2020){Gravity Collaboration},
  {Abuter}, {Amorim}, {Baub{\"o}ck}, {Berger}, {Bonnet}, {Brandner}, {Cardoso},
  {Cl{\'e}net}, {de Zeeuw}, {Dexter}, {Eckart}, {Eisenhauer}, {F{\"o}rster
  Schreiber}, {Garcia}, {Gao}, {Gendron}, {Genzel}, {Gillessen}, {Habibi},
  {Haubois}, {Henning}, {Hippler}, {Horrobin}, {Jim{\'e}nez-Rosales}, {Jochum},
  {Jocou}, {Kaufer}, {Kervella}, {Lacour}, {Lapeyr{\`e}re}, {Le Bouquin},
  {L{\'e}na}, {Nowak}, {Ott}, {Paumard}, {Perraut}, {Perrin}, {Pfuhl},
  {Rodr{\'\i}guez-Coira}, {Shangguan}, {Scheithauer}, {Stadler}, {Straub},
  {Straubmeier}, {Sturm}, {Tacconi}, {Vincent}, {von Fellenberg}, {Waisberg},
  {Widmann}, {Wieprecht}, {Wiezorrek}, {Woillez}, {Yazici}, \&
  {Zins}}]{2020A&A...636L...5G}
{Gravity Collaboration} {et~al.}, 2020, \aap, 636, L5

\bibitem[{{Harko} {et~al}\mbox{.}(2011){Harko}, {Lobo}, {Nojiri}, \&
  {Odintsov}}]{2011PhRvD..84b4020H}
{Harko} T., {Lobo} F. S.~N., {Nojiri} S., {Odintsov} S.~D., 2011, \prd, 84,
  024020

\bibitem[{{Iorio}(2019)}]{2019EPJC...79..816I}
{Iorio} L., 2019, Eur. Phys. J. C, 79, 816

\bibitem[{{Kisslinger} \& {Das}(2019)}]{2019IJMPA..3430013K}
{Kisslinger} L.~S., {Das} D., 2019, Int. J. Mod. Phys. A, 34, 1930013

\bibitem[{{Kopeikin}, {Efroimsky} \& {Kaplan}(2011){Kopeikin}, {Efroimsky}, \&
  {Kaplan}}]{2011rcms.book.....K}
{Kopeikin} S., {Efroimsky} M., {Kaplan} G., 2011, {Relativistic Celestial
  Mechanics of the Solar System}. Weinheim: Wiley-VCH

\bibitem[{{Kramer} {et~al}\mbox{.}(2021){Kramer}, {Stairs}, {Manchester},
  {Wex}, {Deller}, {Coles}, {Ali}, {Burgay}, {Camilo}, {Cognard}, {Damour},
  {Desvignes}, {Ferdman}, {Freire}, {Grondin}, {Guillemot}, {Hobbs}, {Janssen},
  {Karuppusamy}, {Lorimer}, {Lyne}, {McKee}, {McLaughlin}, {M{\"u}nch},
  {Perera}, {Pol}, {Possenti}, {Sarkissian}, {Stappers}, \&
  {Theureau}}]{2021PhRvX..11d1050K}
{Kramer} M. {et~al.}, 2021, Phys. Rev. X, 11, 041050

\bibitem[{{Le Verrier}(1859)}]{LeVer1859}
{Le Verrier} U., 1859, Cr. Hebd. Acad. Sci., 49, 379

\bibitem[{{Lobo}(2009)}]{Lobo09}
{Lobo} F.~S.~N., 2009, in Dark Energy-Current Advances and Ideas, {Choi} J.~R.,
  ed., Research Signpost, pp. 173--204

\bibitem[{{Lucchesi} \& {Peron}(2010)}]{2010PhRvL.105w1103L}
{Lucchesi} D.~M., {Peron} R., 2010, \prl, 105, 231103

\bibitem[{{Lucchesi} \& {Peron}(2014)}]{2014PhRvD..89h2002L}
{Lucchesi} D.~M., {Peron} R., 2014, \prd, 89, 082002

\bibitem[{{Lyne} {et~al}\mbox{.}(2004){Lyne}, {Burgay}, {Kramer}, {Possenti},
  {Manchester}, {Camilo}, {McLaughlin}, {Lorimer}, {D'Amico}, {Joshi},
  {Reynolds}, \& {Freire}}]{2004Sci...303.1153L}
{Lyne} A.~G. {et~al.}, 2004, Science, 303, 1153

\bibitem[{{Milani}, {Nobili} \& {Farinella}(1987){Milani}, {Nobili}, \&
  {Farinella}}]{Nobilibook87}
{Milani} A., {Nobili} A., {Farinella} P., 1987, {Non-gravitational
  perturbations and satellite geodesy}. Adam Hilger, Bristol

\bibitem[{{Nojiri} \& {Odintsov}(2007)}]{2007IJGMM..04..115N}
{Nojiri} S., {Odintsov} S.~D., 2007, International Journal of Geometric Methods
  in Modern Physics, 04, 115

\bibitem[{{Nojiri}, {Odintsov} \& {Oikonomou}(2017){Nojiri}, {Odintsov}, \&
  {Oikonomou}}]{2017PhR...692....1N}
{Nojiri} S., {Odintsov} S.~D., {Oikonomou} V.~K., 2017, \physrep, 692, 1

\bibitem[{{Shapiro}(1990)}]{1990grg..conf..313S}
{Shapiro} I.~I., 1990, in General Relativity and Gravitation, 1989, {Ashby} N.,
  {Bartlett} D.~F., {Wyss} W., eds., Cambridge University Press, Cambridge, pp.
  313--330

\bibitem[{{Shapiro}, {Ash} \& {Smith}(1968){Shapiro}, {Ash}, \&
  {Smith}}]{1968PhRvL..20.1517S}
{Shapiro} I.~I., {Ash} M.~E., {Smith} W.~B., 1968, Phys. Rev. Lett., 20, 1517

\bibitem[{{Shapiro} {et~al}\mbox{.}(1972){Shapiro}, {Pettengill}, {Ash},
  {Ingalls}, {Campbell}, \& {Dyce}}]{1972PhRvL..28.1594S}
{Shapiro} I.~I., {Pettengill} G.~H., {Ash} M.~E., {Ingalls} R.~P., {Campbell}
  D.~B., {Dyce} R.~B., 1972, Phys. Rev. Lett., 28, 1594

\bibitem[{{Shapiro} {et~al}\mbox{.}(1971){Shapiro}, {Smith}, {Ash}, \&
  {Herrick}}]{1971AJ.....76..588S}
{Shapiro} I.~I., {Smith} W.~B., {Ash} M.~E., {Herrick} S., 1971, Astron. J.,
  76, 588

\bibitem[{{Soffel}(1989)}]{Sof89}
{Soffel} M.~H., 1989, Relativity in Astrometry, Celestial Mechanics and
  Geodesy. Springer, Heidelberg

\bibitem[{{Soffel} \& {Han}(2019)}]{SoffelHan19}
{Soffel} M.~H., {Han} W.-B., 2019, {Applied General Relativity}, {Astronomy and
  Astrophysics Library}. Springer Nature Switzerland, Cham

\bibitem[{{Wechsler} \& {Tinker}(2018)}]{2018ARA&A..56..435W}
{Wechsler} R.~H., {Tinker} J.~L., 2018, \araa, 56, 435

\end{thebibliography}

\end{document}